\documentclass[
    ,final            
  ]
  {aipproc}

\layoutstyle{6x9}

\begin{document}

\title{Neutrino Astronomy at the South Pole:\\latest Results from AMANDA-II}


\author{Paolo Desiati, for the IceCube Collaboration\footnote{For the Collaboration list see {http://www.icecube.wisc.edu/pub\_and\_doc/conferences/panic05/}}}{
  address={University of Wisconsin, Madison, WI, U.S.A.}
}



\begin{abstract}
AMANDA-II is the largest neutrino telescope collecting data at the moment,
and its main goal is to search for sources of high energy extra-terrestrial
neutrinos. The detection of such sources could give non-controversial evidence
for the acceleration of charged hadrons in cosmic objects like Supernova
Remnants, Micro-quasars, Active Galactic Nuclei or Gamma Ray Bursts. No significant
excess has been found in searching for neutrinos from both point-like and non-localized
sources. However AMANDA-II has significantly improved analysis techniques for
better signal-to-noise optimization. The km$^3$-scale IceCube telescope will enlarge the observable energy range
and improve the sensitivities of high energy neutrino searches due to its 30 times
larger effective area.
\end{abstract}

\maketitle


\section{Introduction}
\label{s:intro}

The detection of extra-terrestrial multi-TeV neutrinos would significantly contribute
to the understanding of the origin,
acceleration and propagation of high energy cosmic rays. It would also influence the
interpretation of results of the most recent gamma ray experiment results, like those of H.E.S.S. \cite{hess}.

Neutrinos can propagate through the Universe without being affected by interactions or
magnetic fields. Therefore they are ideal high energy cosmic messengers and, for
the same reason, unfortunately very difficult to detect. Even the so called "guaranteed" neutrino fluxes,
i.e. those associated with known cosmic accelerators with an identified pion production target, typically require a $\sim$
km$^3$ scale detector. AMANDA-II \cite{amanda,amanda2}, with its $\sim 0.016$ km$^3$ instrumented volume,
is currently the largest operating neutrino telescope. Its sensitivity is improving because of the larger
data sample collected, better apparatus understanding, and more efficient background rejection
techniques.

The main background for neutrino telescopes is the intense flux of downward penetrating cosmic
muons which can mimic the upward neutrino-induced event signature. A total rejection factor of
$10^6$ is required to isolate muons arising from neutrinos created in atmospheric cosmic ray showers. AMANDA-II has
improved reconstruction algorithms~\cite{reco} and analysis techniques
and currently it is able to select about four upward going muons per day
\cite{icrc05}. After rejecting the downward muon background, these atmospheric neutrinos remain as an irreducible
background in searches for extraterrestrial neutrino fluxes.

The following sections describe the searches for localized neutrino sources and for a diffuse
neutrino flux from many weak sources, respectively.

\section{Neutrinos from point-like sources}
\label{s:point}

\begin{figure}[h]
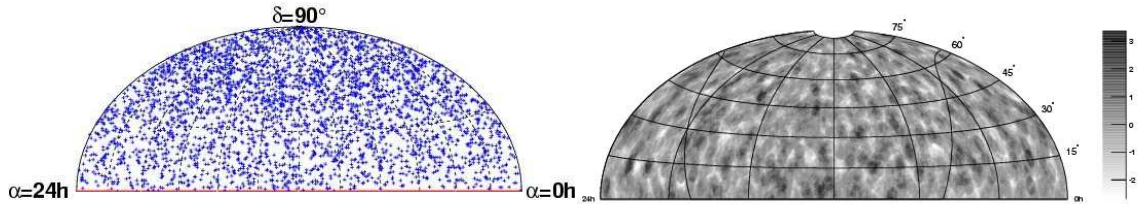

  \includegraphics[height=.12\textheight]{Bernardini_Fig1.epsi}
  \includegraphics[height=.11\textheight]{Bernardini_Fig2.epsi}
  \caption{Left: sky map of the selected 3329 neutrino events from the year 2000-03. Right: significance map
           of the search for clusters of events in the northern hemisphere, based on the selected events (gray
           scale is in sigma).}
  \label{fig:point}
\end{figure}

Since the main background consists of downward-going cosmic ray muons, the search for neutrinos is restricted
to the northern hemisphere. A point source in the sky would be identified as a 
localized significant excess of events, above the irreducible measured atmospheric muon neutrino background.
The event selection is independently performed on different declination bands, for assumed spectra with
$E^{-2}$ to $E^{-3}$ energy dependence. The band width depends on the angular resolution at the given declination.
The optimization of event selection is done by requiring the
best sensitivity, i.e. the best 90\% CL expected average upper limit in the case of no signal \cite{mrf},
at each declination band. The final sample contains 3329 events in the first 807 days of lifetime of AMANDA-II
(see Fig \ref{fig:point}), which has an estimated cosmic ray muon background contamination of less than 10\% and corresponds to a median
angular resolution of $\sim 1.5^{\circ} - 2.5^{\circ}$, depending of declination.
The sensitivity, for an assumed energy spectrum of $E^{-2}$, is $6\times 10^{-8}$ GeVcm$^{-2}$s$^{-1}$, and it is almost independent on declination
\cite{amanda,icrc05}. No excess with respect to the expected background has been observed so far \cite{point1,point2}.

The absence of a significant excess could simply be caused by the small size of the AMANDA-II apparatus and faint
neutrino fluxes from the point sources. Besides increasing the collected data sample, a possibility for improving the signal-to-noise
ratio is to search for events from a pre-selected list of source candidates \cite{amanda2,point1,point2}, or for events in
coincidence with known periods of enhanced electromagnetic emission of selected objects \cite{amanda2}. Another possibility
is to perform a stacking source search using well defined classes of Active Galactic Nuclei, under the assumption that
neutrino emission is similar to the electromagnetic one
\cite{stacking}. With respect to the single point source candidate search the stacking analysis achieves $\sim$ 2-3 times better sensitivity
for some of the selected source classes.

Searches for neutrinos in coincidence with Gamma Ray Bursts were also performed~\cite{mike}, where the time constraint significantly
reduces the background contamination, but no signal was found.

\section{Neutrinos from non-localized sources}
\label{s:diffuse}

The search for high energy neutrinos from all the possible sources in the Universe emitted throughout its evolution
could enhance a very faint signal not detectable from single sources. The search for a diffuse neutrino flux relies on the
simulation of the background and the signal-to-noise optimization is performed on the energy estimation of the selected
events. It is expected that neutrinos have a hard energy spectrum proportional to $E^{-2}$, whereas the atmospheric neutrinos
have a steeper spectrum ($\sim E^{-3.7}$). Therefore, after having assured the quality of selected events, an energy cut is
optimized to have the best sensitivity. The preliminary sensitivity for the search of diffuse muon neutrinos is
$9.5\times 10^{-8}$ GeVcm$^{-2}$s$^{-1}$sr$^{-1}$ for 807 days of livetime, in the energy range between 13 TeV and 3.2 PeV
\footnote{which contains 90\% of the selected events}. This preliminary search did not reveal any significant excess with
respect to the background.

A search for all neutrino flavor events undergoing neutral current interaction as well was also performed. Reconstruction of the
cascade position and energy allows to extend the search to the full sky (not only the northern hemisphere), since 
shower events could be discriminated from tracks. Even so, the major background consists of bremsstrahlung showers produced by cosmic muons
\cite{cascades}. At Ultra High Energy ranges (i.e. above 10$^3$ TeV) the events are very extensive and full reconstruction
is not necessary. At these energies the background of atmospheric events starts to become negligible and the search
can be done by counting the fraction of optical sensors with more than one detected photon, which increases with energy
\cite{uhe}.

The absence of signal detection in AMANDA-II is an indication that the flux of extra-terrestrial neutrinos,
if any, is very small. A $\sim$ km$^3$-scale experiment may open this new observation window by profiting not only of
its larger size and more powerful analysis techniques being developed in AMANDA-II, but also of the full digitized
response of IceCube array sensors \cite{spencer}.

\end{document}